\documentclass[dvips,11pt]{article}
\usepackage{graphicx}
\begin{document}
\title{\bf An Approach to Explain the Long Cooling Times of Anomalous X-ray Pulsars and Soft Gamma Repeaters with the Neutrino Magnetic Moment}

\author{Efe Yazgan\thanks{Corresponding author: efyazgan@metu.edu.tr}~    and Mehmet T. Zeyrek
\\ \\
Department of Physics, Middle East Technical University, 06531, Ankara, Turkey}

\date{}
\maketitle
\begin{abstract}
The surface temperatures as well as X-ray luminosities of Anomalous X-ray Pulsars and Soft Gamma Repeaters are several times higher than that of the ordinary isolated neutron stars at similar ages. We present a simple approach to  explain this observation by the effect of the neutrino magnetic moment on neutron star cooling. This requires a neutrino magnetic moment about $(3-5)\times10^{-11}\mu_B$. The simple approach presented in this paper might be used in determining the exact value of the neutrino magnetic moment and its contribution to the neutrino-electron scattering when the models to explain Anomalous X-ray Pulsars and Soft Gamma Repeaters are mature enough.  

\end{abstract}

\section{Introduction}
In the last few years, strong evidence of neutrino oscillations is obtained implying that neutrinos definitely have mass$^{1-7}$. The existence of mass makes it almost certain that neutrinos also have magnetic moment. The properties of the magnetic moment depends on physics beyond the standard model. The theoretical estimation$^{8,9}$ from the minimally-extended standard model with massive Dirac neutrinos is $\mu_\nu\sim10^{-19}$ in terms of Bohr magneton $\mu_B=e/2m$. This value is much smaller than the current experimental upper-limit$^{10-13}$ $\mu_\nu<1.0-1.3\times10^{-10}\mu_B$. It is important to note that the theoretically predicted value can be increased to the upper-limit value by including the effects of Majarona transition moments, right-handed currents, supersymmetry, extradimensions or other physical models beyond the standard model $^{14-18}$. 
There might be many astrophysical effects of a non-zero neutrino magnetic moment. 
Although there are several other neutrino processes, neutron stars in the temperature interval from $\sim10^9$ K to $\sim10^6$ K predominantly cool by $\beta$-processes with additional neutrons to conserve energy and momentum in the degenerate interior$^{19}$ $n+n\rightarrow p+n+e^{-}+\bar{\nu_e}$ and $n+p+e^{-}\rightarrow n+n+\nu_e$.  This temperature interval corresponds to few days to about $10^5$ years after the neutron star is born. When the temperature falls below $\sim10^6$ K, photon emission from the surface gradually begins to dominate. 
The photon cooling period is considerably shorter than the neutrino cooling period. Thus, neutrino emission is the main cooling agent for neutron stars with ages less than $10^6$ years $^{20}$. 
Isolated neutron stars have high magnetic fields, therefore the non-zero neutrino magnetic moment might play a significant role in neutron star cooling and might lead to some observational consequences for these sources.

\section{Surface Temperature of Isolated Neutron Stars, Anomalous X-ray Pulsars and Soft Gamma Repeaters}
Ordinary isolated neutron stars have magnetic fields about $10^{12}$ G and until they are $\sim10^5$ years old, they primarily cool down by neutrino emission $^{19,21,22}$. 
In the last decade, a new class of isolated neutron stars were identified. This class includes anomalous X-ray pulsars (AXPs) and soft gamma repeaters (SGRs) $^{23}$. It is believed that AXPs and SGRs have surface dipole magnetic fields$^{24}$ of $10^{14}-10^{15}$ G. A significant observation is that AXPs and SGRs do not cool as fast as ordinary isolated neutron stars $^{25-26}$. 
Figure 1 displays the surface temperatures, or equivalently blackbody temperatures, of AXPs/SGRs$^{27}$ and ordinary isolated neutron stars$^{28}$ as a function of age. In the figure '+' sign represents AXPs and SGRs and 'x' sign represents ordinary isolated neutron stars. It must be stressed that for the ordinary neutron stars younger than $10^5$ years, we have used the surface temperatures estimated from modeling with light-element atmospheres, because only these models give results consistent with the canonical values for neutron star radii $^{28-30}$. It is seen from the figure that the ratio of the temperatures of AXPs/SGRs and ordinary isolated neutron stars ($T_{as}/T_{ns}$) at similar ages ranges from $\sim$2.5 to 8. 
The difference in cooling times manifests itself also on the X-ray luminosities of these objects. These observations might be explained by a $\sim10^{15}$ G magnetic dipole field which is decaying in time (Magnetar Model)$^{24}$ or a lighter neutron star possessing a magnetic dipole field of $\leq10^{14}$ G $^{26,35}$. 
Another alternative explanation for the high X-ray luminosities is that the neutron star might have a magnetic field of $\sim10^{15}$ G and it possesses a thin insulating envelope of matter of low atomic weight at densities$^{36}$ $\rho<10^7-10^8$ g/cm$^3$. As another alternative, we speculate that this difference, at least partially, is a result of a finite neutrino magnetic  moment interacting with the very high magnetic fields of AXPs and SGRs which is $\sim10^{14}-10^{15}~G$. In the next section, we find the value of the neutrino magnetic moment to account for the difference of surface temperatures by assuming that the difference in the cooling times is solely a result of the interaction of the neutrino magnetic moment with the neutron star magnetic field.

\subsection{Spin-Flip Precession and Cooling}
The following transformations are possible when a left handed electron neutrino encounters strong magnetic fields$^{37}$; $\nu_{e_L}\rightarrow\nu_{e_R}$, $\nu_{e_L}\rightarrow\nu_{\mu_R}$ and $\nu_{e_L}\rightarrow\overline{\nu}_{\mu}$ .
The first two conversions refer to Dirac neutrinos and the last conversion refers to Majorana neutrinos. Likhachev and Studenikin$^{37}$ derived a value for the critical magnetic field above which the neutrino oscillations become important. Their critical magnetic field depends on the mass squared difference, mixing angle, effective matter density, neutrino energy and magnetic moment. They have estimated, taking the neutrino magnetic moment $10^{-10}\mu_B$ and reasonable neutron star parameters the critical magnetic field to be 
\begin{equation}
B_{cr}\sim\alpha\equiv\sqrt{2}G_Fn_{eff}\sim10^{14}~G 
\end{equation}
This corresponds to an effective oscillation length of 
\begin{equation}
L\sim2\pi\left[\alpha^2+(2\mu B_{cr})^2\right]^{-1/2}\sim1~cm
\end{equation}
which is 6 orders of magnitude smaller than the radius of the neutron star.
This critical field is calculated without taking into account the rotation of the neutron star which might cause the neutrino to be affected by a time-dependent magnetic field in a plane transverse to the neutrino trajectory as it propagates from the core to the surface. Taking the magnetic field variations can reduce the critical magnetic field. 
For a neutron star, the critical magnetic field is determined from the matter term $\alpha$. Therefore, the critical field will decrease and thus the oscillation probability will increase toward the surface. 

What cools the neutron star is neutrino production. The neutrinos can reheat the neutron star by inelastic  scatterings with electrons via $\nu_e+e^-\rightarrow\nu_e+e^-$. There are other reactions which are also important for reheating such as neutrino-nucleon absorption and antineutrino-proton absorption, but their cross-sections are similar to neutrino-electron scattering and makes the mean free path only several times smaller$^{21}$. The mean free path for neutrino-electron scattering process$^{21}$ is $\sim9\times10^7~km$. However, neutrinos are also scattered {\it elastically} by neutrons. Elastic scattering prevents the neutrino to escape from the neutron star directly. This increases the probability for inelastic scattering from electrons. The modified mean free path is given as$^{21}$
\begin{equation}
\lambda\sim(\lambda_n\lambda_e)^{1/2}\sim2\times10^{5}~km
\end{equation}
 where $\lambda_n$ and $\lambda_e$ corresponds to the mean free path for neutron and electron scattering respectively.  
However, this is still much higher than the neutron star radius. We speculate that the interaction of the neutrino magnetic moment with the magnetic field is a scattering process that the effective oscillation length given in Eqn. [2] corresponds to mean-free path for magnetic scattering. By analogy, the total mean free path is then
\begin{equation}
\lambda_{tot}\sim(\lambda L)^{1/2}\sim1.4~km
\end{equation}
This is an order of magnitude smaller than the neutron star radius. We would like to point out that the value above is calculated by assuming $B\sim10^{14}~G$. However, this is a value even smaller than the interior magnetic fields of ordinary neutron stars. It is believed that the surface magnetic field of AXPs/SGRs are $10^{14}-10^{15}~G$ and just above the core it can be as high as $\sim10^{17}~G$ by flux conservation. Thus the mean free path calculated above is the maximum mean free path. 

From the above arguments we can calculate an upper limit for the neutrino magnetic moment from Figure 1. If the difference in the surface temperatures is solely explained by neutrino magnetic moment then the ratios of effective oscillation lengths must be inversely proportional with the temperatures. Therefore,
\begin{equation}
\frac{L_{as}}{L_{ns}}=\sqrt{\frac{\alpha^2+(2\mu B_{ns})^2}{\alpha^2+(2\mu B_{as})^2}}=\frac{T_{ns}}{T_{as}}
\end{equation}
The ratio ($T_{as}/T_{ns}$) at similar ages are $\sim$2.5-8, $B_{as}=10^{15}~G$ and $B_{ns}=10^{12}~G$. Plugging these values in the above equation gives
$\mu\approx(3-5)\times10^{-11}\mu_B$.

Instead of surface temperatures we could utilize the X-ray luminosities.
For the thin atmosphere model, the X-ray luminosities of AXPs and SGRs can be determined$^{36}$. We have found that the ratio of observed to the predicted values of X-ray luminosity $(L_{obs}/L_{pre})$ changes from 1 to 6.5. Only for AXP 1E1048.1-5937 the above ratio is smaller than 1. Considering the errors in the observed X-ray luminosity, we obtain $L_{obs}/L_{pre}\sim0.5-13$. A neutrino magnetic moment determined from this ratio of luminosities is consistent with the value we have determined from the ratio of surface temperatures.

\section{Discussion and Conclusion}
The estimation above depends on the crucial assumption that the neutrino scattering mean free path is at least a fraction of the effective oscillation length. This is possible since the total cross-section for neutrino-electron scattering depends on the oscillation probability$^{38}$. However, the exact dependence of the cross-section on the oscillation probability is not known and depends on the physics beyond the standard model. Future observations of AXPs and SGRs will constrain the models and will provide valuable information not only on the value of the magnetic moment, but also on the dependences of mean free paths and cross sections on the magnetic moment. But from our current knowledge of AXP and SGR astronomy we can only conclude that the neutrino magnetic moment value quoted above is an upper-limit.

$Acknowledgments$.
It is a pleasure to thank O.H. Guseinov, A. Ankay and H. Gams\i zkan for fruitful and stimulating
discussions.

\begin{figure}[h]
\includegraphics[width=9cm,angle=-90]{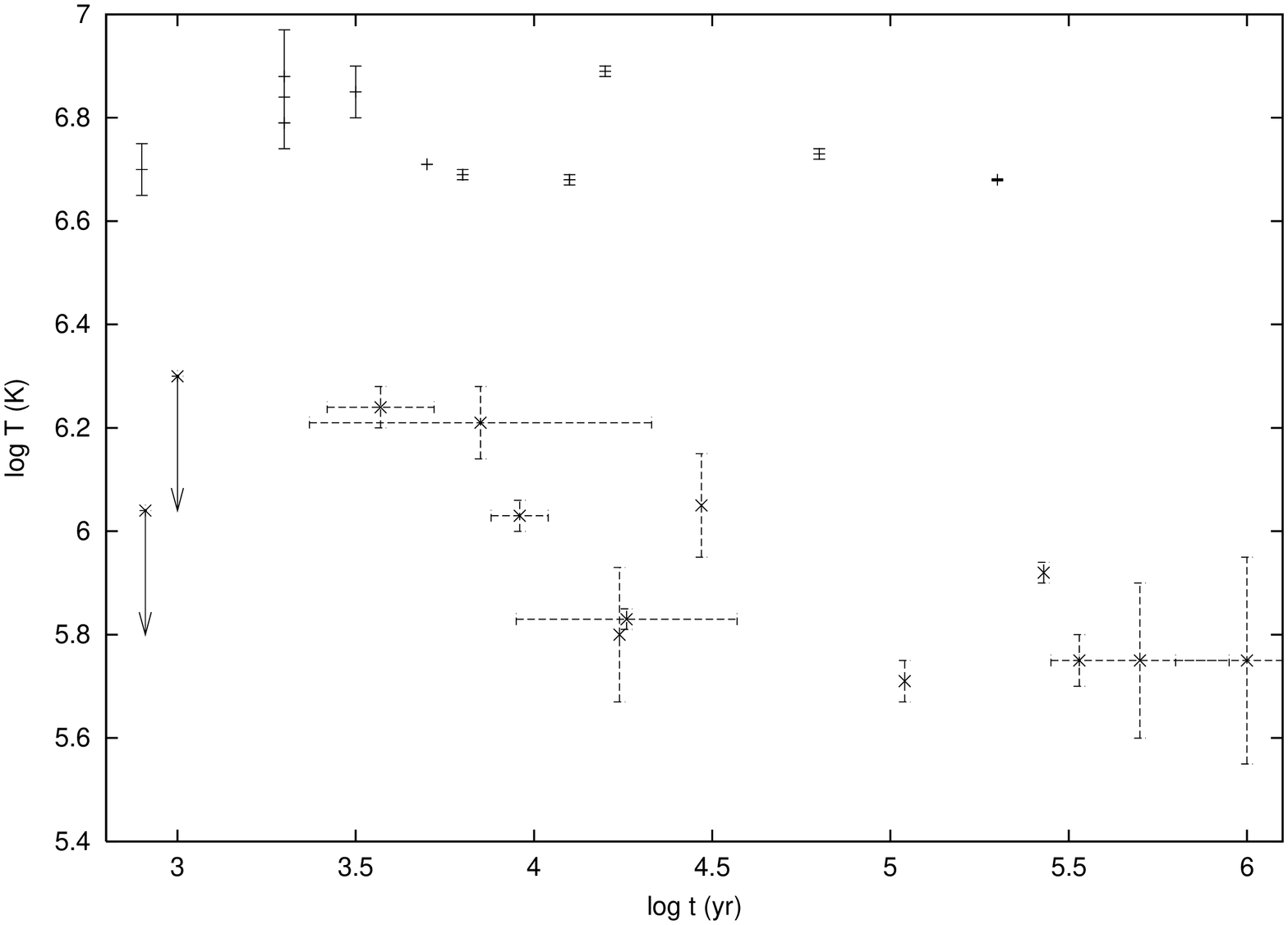}
\caption{Thermal histories of AXPs/SGRs and ordinary neutron stars.'+' represents AXP/SGR and 'x' represents ordinary neutron stars.}
\end{figure}

\end{document}